\documentclass[tradit]{aa}

\usepackage{amsmath}
\usepackage{natbib, twoopt}

\usepackage{graphicx}
\usepackage{color}
\usepackage{hyperref}
\usepackage{breakurl}

\usepackage{soul}
\sethlcolor{yellow}

\usepackage{txfonts}

\makeatletter
\newcommand\tinyv{\@setfontsize\tinyv{4pt}{6}}
\renewcommand*{\@fnsymbol}[1]{\ifcase#1\or*\or$\dagger$\or$\ddagger$\or**\or$\dagger\dagger$\or$\ddagger\ddagger$\fi}
\makeatother

\definecolor{caribbeangreen}{rgb}{0.8, 0.6, 0.79}

\definecolor{blizzardblue}{rgb}{0.95, 0.65, 1.0}

\definecolor{lightblue}{rgb}{.90,.90,1}
\definecolor{lightorange}{rgb}{0.996, 0.847, 0.694}
\definecolor{lightgreen}{rgb}{0.82,0.94,0.75}
\definecolor{lightviolet}{rgb}{0.86, 0.82, 1}
\definecolor{turq}{rgb}{0.68, 0.89, 0.87}
\definecolor{blond}{rgb}{0.98, 0.94, 0.75}
\definecolor{cherryblossompink}{rgb}{1.0, 0.72, 0.77}

\definecolor{palegreen}{rgb}{0.82,0.94,0.75}
\definecolor{brilliantlavender}{rgb}{0.96, 0.73, 1.0}

\definecolor{mistyrose}{rgb}{1.0,0.894,0.882}

\definecolor{bananamania}{rgb}{0.98, 0.91, 0.71}

\newcommand{\hlru}[2][white]{{\sethlcolor{#1}\hl{#2}}}

\newcommand{\hlr}[2][white]{{\sethlcolor{#1}\hl{#2}}}

\newcommand{\hdm}[2][white]{{\sethlcolor{#1}\hl{#2}}}

\newcommand{\hlw}[2][white]{{\sethlcolor{#1}\hl{#2}}}

\newcommand{\hlm}[2][white]{{\sethlcolor{#1}\hl{#2}}}

\newcommand{\hld}[2][white]{{\sethlcolor{#1}\hl{#2}}}
\newcommand{\hll}[2][white]{{\sethlcolor{#1}\hl{#2}}}

\newcommand{\hlz}[2][white]{{\sethlcolor{#1}\hl{#2}}}

\newcommand{\hlv}[2][white]{{\sethlcolor{#1}\hl{#2}}}
\newcommand{\hlo}[2][white]{{\sethlcolor{#1}\hl{#2}}}

\newcommand{\hbb}[2][white]{{\sethlcolor{#1}\hl{#2}}}

\newcommand{\fermilat}{{\it Fermi}-LAT}
\newcommand{\gray}{$\gamma$-ray}
\newcommand{\grays}{$\gamma$ rays}

\long\def\LEt#1{}

\def\apj{ApJ}                 
\def\apjl{ApJ}                
\def\apjs{ApJS}               
\def\aap{A\&A}                
\def\mnras{MNRAS}             
\def\prl{Phys.~Rev.~Lett.}    
\def\prd{Phys.~Rev.~D.}

\makeatletter
\renewcommand*{\@fnsymbol}[1]{\ifcase#1\or*\or$\dagger$\or$\ddagger$\or**\or$\dagger\dagger$\or$\ddagger\ddagger$\fi}
\makeatother

\begin{document}

\titlerunning{Search for intraday transients}
\authorrunning{D. A. Prokhorov \& A. Moraghan}

\title{A blind search for intraday gamma-ray transients with \textit{Fermi}-LAT: Detections of GRB and solar emissions}

\author{D. A. Prokhorov\inst{1,2}\thanks{E-mail:phdmitry@gmail.com}
  \and A. Moraghan\inst{3,4}\thanks{E-mail:anthony.moraghan@manchester.ac.uk}}

\institute{GRAPPA, Anton Pannekoek Institute for Astronomy, University of Amsterdam, Science Park 904, 1098 XH Amsterdam, The Netherlands
  \and Institute of Physics, Academia Sinica, Taipei, 11529, Taiwan
  \and Academia Sinica Institute of Astronomy and Astrophysics, 11F
of AS/NTU Astronomy-Mathematics Building, No.1, Sec. 4, Roosevelt Rd.,
Taipei 10617, Taiwan
  \and Jodrell Bank Centre for Astrophysics, University of Manchester, Manchester M13 9PL, UK}

\date{\today}



\abstract{
\LEt{***General notes. a) I have edited to UK English convention as this most closely matches your spelling and punctuation use. b) A\&A uses the past tense to describe the specific steps used in a paper and the present tense to describe general methods and recent findings. Please make sure this is followed throughout the paper. For details, see Sect. 6 of the Language Guide: https://www.aanda.org/for-authors/language-editing/6-verb-tenses  c) All abbreviations and acronyms are introduced at first use, once in the Abstract and again in the main text. Instrument and program names are introduced at first use (when appropriate) in the main text. All abbreviations should be used consistently. Please check throughout }\\We present a search for intraday transient $\gamma$-ray signals using \hlr{15.4} years of the \textit{Fermi} Large Area Telescope data. The search is based on a recently developed variable-size sliding-time-window (VSSTW) analysis and is aimed at studying variable $\gamma$-ray emission from gamma-ray bursts (GRBs) and the Sun. We refined the algorithm for searches for transient sources in order to solve the search problem within a reasonable amount of CPU time. These refinements allowed us to \hbb{increase} the number of gamma-ray bursts, solar flares, and quiescent solar events detected with the VSSTW technique by several times compared to the previous VSSTW search. \hbb{The current} search revealed a new $\gamma$-ray \hdm{signal} recorded with \textit{Fermi}-LAT on 2018 January 12. This \hdm{signal is probably from} a GRB and deserves an exploration of the existing archival multiwavelength observations in order to identify it in an unambiguous way. \hdm{We also report a $\gamma$-ray signal from the solar flare on 2023 December 31 that occurred during the 25th solar cycle.}}


\maketitle

\makeatletter
\renewcommand*{\@fnsymbol}[1]{\ifcase#1\@arabic{#1}\fi}
\makeatother

\renewcommand{\labelitemi}{$\bullet$}

\section{Introduction}
\label{Sect1}

\hlz{Transient sources are related to explosive and catastrophic events where very compact objects are involved and usually imply extreme astrophysical environments. The difference between a variable object and a transient object can be defined from the observational point of view on the detection limit of the instrument: a variable object remains detectable at any time, while a transient source does not. Owing to their serendipitous occurrence, time variability, and duration on different timescales,}
\citet[][]{Akerlof2011} used a good phrase to describe searches for \hlz{some} unrepresentative
$\gamma$-ray transients: {Not every fish can be caught with the same net}.\LEt{***direct quotes are discouraged (and likely Akerlof did not invent this sentence, so he shouldn't be cited. Does this work?) Also: A\&A does not use italics for emphasis. Please remove this formatting here and throughout your paper. Italics for instrument names can remain  } The term `blind' (or unguided) used in the paper title \hlv{emphasises} that this type of search has no additional information about states beyond that provided in the problem definition. In the two previous papers \citep[][]{PMV2021, PM2023}, we \hlv{binned $\gamma$ rays into} time intervals of one week \hlv{and performed temporal studies using this selection of intervals}. In the \textit{Fermi} All-sky Variability Analysis \citep[FAVA;][]{FAVA1, FAVA2}, the authors also selected one-week time intervals. The one-week time intervals are \hlo{basic to} \textit{Fermi}-LAT \hlz{data as it is} \hlo{provided} in weekly data files and \hlo{are} suitable for searches for transients lasting for several weeks and for short, but \hlz{brighter} transients. For instance, many \gray{} bursts (GRBs) last for less than a few hours, and if six-hour time intervals are used instead of one-week time intervals, the gain in a signal-to-noise ratio (S/N) \LEt{***Please use S/N to abbreviate signal-to-noise ratio. A\&A uses SNR for supernova remnant. I have changed them in the text and captions, but please check within the tables and figures themselves} compared to one-week time intervals for a short, hour-long $\gamma$-ray transient is $\sqrt{7\times24/6}\approx 5.3$, where $7\times 24$ and 6 correspond to the two intervals in   units of hours. This \hld{gain} is due to a \hld{smaller} number of background \grays{} hitting a $\gamma$-ray detector over a \hld{shorter} period of time. Since the factor of 5.3 is fairly large,
we can expect   detections of short \gray{} transients \hbb{that remained below the detection thresholds} in the previous blind searches \hlz{for transients}.\LEt{or perhaps: we can expect that detections of trasients  will remain below}

The \hbb{\texttt{PYTHON}} code, which employs a likelihood analysis  to find the most statistically significant time interval of a high flux at a given position in the sky, developed by \citet[][]{PMV2021}, is publicly available on Zenodo\footnote{\burl{https://zenodo.org/record/4739389}} \citep[for
a review of likelihood theory, see][]{Mattox}. Following this approach, we performed a \hlv{new} search for transient \gray{} sources and, for this purpose, we compared a model with the presence of a temporary bright state above a steady flux level and a model assuming a source with a steady flux in time. In the previous papers, the variable-size sliding-time-window (VSSTW) approach was used \hlo{particularly} to identify bright $\gamma$-ray flux \hlo{transient events} \hlz{each} encompassing many consecutive \hlo{weekly} intervals. \hlv{Here} we used the VSSTW approach to \hld{find} short transient events
\hld{confined within single time intervals}, such as \hlv{most of} \hlz{the} GRBs and solar emissions, and to distinguish \hld{these events} from ubiquitous flares of active galactic nuclei (AGNs) which \hld{typically} last for a longer time. The number of sliding window time intervals increases with a decreasing time step as $\mathcal{O}(N^{2}$), where $N$ is the total number of intervals. Therefore, the computing time increases quadratically and will be approximately $(7\times24/6)^2\approx1000$ times longer, making such a search computationally expensive. Since the previous search \hlv{by} \citet[][]{PMV2021} took about 16000 CPU hours, we revised the algorithm \hld{in order to solve this search} problem \hlv{within a reasonable amount of} \hlo{CPU time}.

\hlr{A total of} 38 transient \hlv{non-AGN} \gray{} signals are \hlo{present} in the \hlv{list} obtained from a VSSTW analysis \hlz{by} \citet[][]{PMV2021}. Among \hlo{these signals}, nine correspond to GRBs, three are  solar flares, and \hbb{two  are solar events due to quiescent solar emission}. \hlo{In the VSSTW analysis,} the Sun, \hlo{which} appears to move across the sky \hlv{by about 1$^{\circ}$ per day}, can be considered  a $\gamma$-ray source moving along the ecliptic over the course of a year. The second \textit{Fermi}-LAT GRB  catalogue \citep[][]{LATgrbCat} and the first \textit{Fermi}-LAT solar flare catalogue \citep[][]{LATsolarCat} contain 186 GRBs and 45 solar flares, respectively. The \hlv{large} difference in numbers of GRBs and solar flares between the list obtained from the \hlz{previous} VSSTW analysis and these two \hbb{catalogues} \hlz{are investigated in this paper by applying more sensitive techniques}. A similar problem of increasing the number of binaries \hld{obtained} from the VSSTW analysis was addressed by \citet[][]{PM2023}, \hlo{who included} $\gamma$ rays with lower energies and \hlo{increased} a number of test positions compared to the previous analysis. However, it is worth mentioning that   blind searches are commonly less sensitive than   searches \hlv{that make use of} multiwavelength information. The spatial and temporal \hld{characteristics} of \hld{multiwavelength emission} from  transient sources are useful for detecting and identifying such signals \hlo{in $\gamma$ rays} \citep[see][]{Akerlof2010, Akerlof2011, Rubtsov2012}. On the one hand, \hlo{the knowlegde of} \hld{this information leads to a small} number of trials compared to blind searches \hlo{and, in turn, to a higher detection significance}. On the other hand, blind searches in the \gray{} band allow detections of signals, which for some reasons are \hlo{missing or} poorly spatially \hlw{localised} \hlz{at other wavelengths}.

We performed a new search \hld{in order to increase the number of transient $\gamma$-ray sources lasting for less than six hours}, in particular GRBs and solar flares, detected with the VSSTW method. The significant gain in  S/N also enabled us to \hlz{search for} new transient $\gamma$-ray sources and \hlr{visualise} \hlw{variations in the flux of $\gamma$ rays from the Sun} across one and a third solar cycles.

\section{Spatiotemporal search method}

In this section we provide an overview of the existing spatiotemporal search methods and describe a fast algorithm for a VSSTW search for short $\gamma$-ray transients.

\subsection{Overview of spatiotemporal search methods}
\label{CA}

The \textit{Fermi} Flare Advocate program, also known as Gamma-ray Sky Watcher (GSW),  provides a quick look and review of the $\gamma$-ray sky observed daily by the \textit{Fermi}-LAT through on-duty LAT Flare Advocates and high-level software pipelines such as the LAT Automatic Science Processing \citep[ASP;][]{Chiang2007} and the \textit{Fermi} All-sky Variability Analysis \citep[][]{FAVA1}. The GSW service provides rapid alerts and communicates to the scientific community potentially new $\gamma$-ray sources, \hlz{remarkable} transients, and flares. News items are regularly posted through Astronomer's Telegrams. However, not all \textit{Fermi}-LAT transient events are reported through Astronomer's Telegrams by the \textit{Fermi}-LAT team; for example,  several high $\gamma$-ray states of Cygnus X-3 were reported by the AGILE team, but not by the \textit{Fermi}-LAT team despite their detections both with AGILE and \textit{Fermi}-LAT \citep[see][]{PM2023}.  It is \hll{a case-by-case decision,} made by on-duty LAT Flare Advocates, \hll{whether or not to} report \hll{a transient or a flare} through an Astronomer's Telegram. \hll{Moreover, the} \textit{Fermi} \hll{gamma-ray sky blog}\footnote{\burl{fermisky.blogspot.com}} \hll{was actively maintained by on-duty LAT Flare Advocates from 2009 to 2017, but has since ceased operations.}

The ASP analysis pipeline, which runs on the photon event and spacecraft data fits files, is composed of several scientific tasks \citep[][]{Chiang2007}. Among these are (i) blind source detection on all-sky photon counts maps accumulated in six-hour and one-day intervals, through a fast sky map scan method based on a two-dimensional Mexican Hat wavelet transform, thresholding, and sliding cell algorithms \citep[][]{Ciprini2007}, and (ii) transient and flare identification based on a variability test. In this paper we investigate the problem particularly related to these two tasks. To the best of our knowledge, the ASP analysis pipeline is available only within the \textit{Fermi}-LAT team. The open source \hbb{code} for a VSSTW search creates new opportunities for members of the $\gamma$-ray astronomy community.

Temporal information plays a crucial role for identification of transient events. Most of the search pipelines for variable $\gamma$-ray emission from \hlz{sources within} the large region of the sky \citep[][]{Neronov12, FAVA1, FAVA2, FermiLTT2021} on a   timescale of weeks or months have one time variable that is the index of a time interval, \hll{which indicates where the high-flux state occurs}. The VSSTW method \citep[][]{PMV2021, PM2023} has two time \hll{variables}, namely the index of a time interval and the duration of a time interval. This second variable \hlw{provides} valuable information, for example for finding \hll{longer} signals, such as high $\gamma$-ray flux states of the $\gamma$-ray binary LS I +61$^{\circ}$303 \hll{present} due to its superorbital period of 55 months. In addition, this variable can also be used to differentiate sources of a given class from \hll{the} other sources on the basis \hll{of} how long their $\gamma$-ray emission typically lasts. For \hlw{most of the} GRBs and solar flares, we expect that the $\gamma$-ray signals lie within one six-hour interval or at most two six-hour intervals (\hll{in the case} of a GRB \hbb{occurring} near a boundary between two six-hour intervals).

Various methods for detecting point-like sources against the photon and instrumental background exist and are  based on the search of local concentrations of $\gamma$ rays. In addition to\LEt{***"besides" is too colloquial for a formal paper. It can also be ambiguous. Depending on context, it can mean "except for" or "in addition to", and so should be avoided. Please check throughout that I have chosen appropriately. } the traditional maximum likelihood algorithm \citep[][]{Mattox}, other methods include the use of wavelet transform
\citep[][]{Damiani1997, Ciprini2007}, the density-based clustering algorithm
DBSCAN \citep[][]{Tramacere2013, Armstrong2015}, and the minimum spanning tree algorithm \citep[][]{Campana2018}. Although these methods \hll{permit searches for} sources within each \hll{given} time interval, the \hll{problem of} how to associate a source \hll{detected} in a particular time interval with a transient remains \hll{unresolved}. To \hll{solve this problem}, it is necessary to take \hbb{temporal information} into account. The VSSTW method, which takes \hbb{both temporal information and spatial information} into account, is an efficient computational framework for searches for $\gamma$-ray transients.

\subsection{VSSTW search method for short transients}
\label{Sect2p2}

The VSSTW method was developed to search for the most statistically significant time interval of a high flux at a given position in\LEt{***on? } the sky by taking  temporal information into account. The time window has a size encompassing a whole number of time intervals and can start from the beginning of \hll{any} of these time intervals. In this paper we show that the VSSTW method provides an \hll{efficient method} of searching for short $\gamma$-ray transients.

The VSSTW method, written in \texttt{PYTHON}, uses data binned  into spatial bins and into time bins. For  spatial binning, the HEALPix package\footnote{\burl{http://healpix.sourceforge.net}} \citep[][]{Gorski} provides a partition of the sky into equal-area pixels. As mentioned \hbb{in Section} \ref{Sect1}, six-hour time bins are suitable for a search for short $\gamma$-ray transients in the \textit{Fermi}-LAT data; however, this data binning requires a faster search algorithm than that used by \citet[][]{PM2023}. With this time binning, there are about 20 thousand six-hour intervals. To accelerate the VSSTW search, we made three important changes.\LEt{***A\&A avoids bullet points and lists where possible (this does not apply to the Conclusion). Here your points are too long and  discursive. Please   reformat as a normal paragraph (or a separate paragraph for each point).  I have made the appropriate changes. }

First, instead of simultaneously opening 20 thousand of count maps and 20 thousand of exposure maps for reading the values for each spatial pixel, we simultaneously opened 1500 count maps and 1500 exposure maps and consequently repeated this process 16 times covering all the 20 thousand intervals. Each of these 16 slices \hlw{approximately} corresponds to one year of data, \hll{except for the first slice and the last slice};  all these slices are independent from each other with respect to a search for short $\gamma$-ray transients. This division helps us to significantly reduce the simultaneous memory (RAM) usage. It should be noted that the first of the 16 slices is shorter and contains 600 six-hour intervals from the start of the \textit{Fermi}-LAT science operation in 2008 August to the end of 2008. Due to this choice, the second slice starts from the beginning of 2009, the third slice starts from the beginning of 2010, and so on. \hdm{The 16th slice is also shorter and contains 944 $\mathrm{six}$-hour intervals of the data accumulated between 2023 May and 2024 January.}

For the second change we computed the flux summed over the \hlr{energy bands used in this paper} for each of the spatial pixels. Given that we searched for short $\gamma$-ray transients for each of the slices, containing 1500 six-hour intervals, we selected 400 six-hour intervals of the highest flux \hlr{for every spatial pixel}. We performed the VSSTW search within and around these selected intervals \hll{using them as `seed intervals'. We note that these seed intervals are expectedly different for each spatial position.}\LEt{***Quote marks should not be used for emphasis. If you are indicating that the word is being used in an unusual way or with a special meaning, the quote marks can remain (`single' quote marks in UK, ``double'' in US), but only for the first use. All subsequent uses of the word should be without quote marks. }

For the third change, compared to \hlw{the searches} performed by \citet[][]{PMV2021} \hlw{and} \citet[][]{PM2023} which cover long transients, the time window can be significantly shortened when searching for short transients. We analysed data belonging to the \hbb{four} time windows starting \hbb{from the beginning} of a seed interval and ending \hbb{within} 24 hours. Thus, the minimum duration of a time window is six hours and the maximum duration of a time window is \hbb{24 hours}.

These three changes allowed \hlr{us} to significantly reduce \hlr{the required CPU time} and avoid the RAM overload. \hlr{We utilised the equivalent of 512 cores on a cluster over a period of 72 hours, resulting in approximately 37000 core-hours of computational time.}

We estimated the number of uncorrelated test spatial positions as $41253/(\pi\times1.5^{2}) = 5836$,
where 41253 deg$^2$ is a full sky area and the radius of $1\fdg5$ is a radius of the PSF 68\% containment at 0.5 GeV and minimises a position correlation of signals. We adapted a
global significance level where we indicated the significance level after taking the `look elsewhere effect' into account. This effect is quantified in terms of a trial factor that is the ratio of the probability of observing the excess in the obtained time interval to the probability of observing it with
the same local significance level anywhere in the allowed range at any of the uncorrelated test positions \citep[e.g.][]{PMV2021}. For this search, we selected two criteria, named strong and weak. The strong criterion requires a global significance level above 5.0$\sigma$ in one of \hlw{the one-year slices}. Given the trial factor, $5836\times1500=8.75\times10^{6}$, the global significance level for the strong criterion translates to a local significance level higher than $7.5\sigma$. The weak criterion requires a global significance level above 3.0$\sigma$ in any of \hlw{the one-year slices}. Given the trial factor, $5836\times1500\times15=1.31\times10^{8}$, the global significance level for the weak criterion translates to a local significance level higher than $6.7\sigma$.

In order to improve the sensitivity for sources whose flaring activity is at the edge of the time intervals, we shifted the data by three-hours and repeated the six-hour binned analysis.

\section{\textit{Fermi}-LAT data}

The \textit{Fermi} Large Area Telescope (LAT) is a pair-conversion \hlz{$\gamma$-ray detector} on board the \textit{Fermi} Gamma-ray Space Telescope (FGST) spacecraft \citep[][]{Atwood2009}. The FGST spacecraft was launched into an initial near-earth orbit of $\simeq$565 km altitude at a  25.6 degree inclination with an eccentricity $<$0.01 on 2008 June 11.\LEt{***see also note 24 and check throughout. Please make sure the date format is consistent throughout the paper (see the Language Editing guide, Sect. 2 Main guidelines for A\&A style, specifically Sect. 2.4. Date Format, https://www.aanda.org/for-authors/language-editing/2-main-guidelines)  } The \textit{Fermi} orbital period is 96.5 minutes. \textit{Fermi}'s orbit keeps it close to the equator, \hld{but intersects the South Atlantic Anomaly region of intense background radiation.}
\textit{Fermi}-LAT \hld{does not collect data when the spacecraft traverses the South Atlantic Anomaly (SAA), which occurs about $\sim$15\% of the time.}
The diameter of the Earth as seen from Fermi is $\simeq135^{\circ}$, so roughly 30\% of the sky is occulted by the Earth at any one time. The field of view of   \textit{Fermi}-LAT covers $\sim$20\% of the sky at any instant \hbb{and up to} 75\% of the sky every orbit. \textit{Fermi}-LAT operates primarily in an all-sky scanning survey mode and \hlz{has been scanning the sky continuously since 2008 August}. \hlz{Until 2018 March, the FGST spacecraft had rocked $50^{\circ}$ north from the orbit plane for one orbit, and then $50^{\circ}$ south from the orbit plane for the next orbit.}
The survey mode affords coverage of the entire sky every three hours. Since 2018 March,  the rocking profile enabling the LAT all-sky survey has been replaced with periods of various alternating rocking angles, but with minimal changes to the \textit{Fermi}-LAT's long-term sky exposure \citep[][]{Fermi10}. \hlz{Given all these temporal scales,} we chose \hlw{six-hour time intervals}, resulting in relatively uniform exposure, for a search for short $\gamma$-ray transients.

\textit{Fermi}-LAT provides an angular resolution per single event of \hlz{$5^{\circ}$ at 0.1 GeV}, $1.5^{\circ}$ at 0.5 GeV, narrowing to $0.8^{\circ}$ at 1 GeV, and further narrowing to $0.1^{\circ}$ above 10 GeV.
At energies below $\sim$10 GeV, the \hld{reconstruction accuracy of $\gamma$-ray arrival directions} is limited by multiple \hld{Coulomb} scattering \hlz{of electron-positron pair components} in the \hlz{heavy-metal} converter foils of \fermilat{}. \hlz{As long as multiple scattering} \hbb{dominates,} the point spread function (PSF) will have a 1/E dependence. Above $\approx$10 GeV, multiple scattering is unimportant and the accuracy is limited by the ratio of silicon-strip pitch to silicon-layer spacing.
Given both the angular resolution dependence with energy and the broadband sensitivity to sources with power-law spectra,\footnote{\burl{http://www.slac.stanford.edu/exp/glast/groups/canda/lat\_Performance.htm}}
we selected the optimal lower energy limit of 0.5 GeV. We made this selection with two purposes: to tighten the PSF and to include $\gamma$ rays with energies \hld{down to} 0.5 GeV, as was done by \citet[][]{PM2023} \hld{to increase the sensitivity for} transient $\gamma$-ray sources \hlz{with soft spectra}, such as Cygnus X-3 \citep[][]{CygX3Sci} and PSR B1259-63 \cite[][]{Tam2011, psrb12592020}. \hld{Gamma-ray} \hlz{emission due to the decay
of neutral pions produced by $>$300 MeV protons and ions, with a power-law spectrum of index between 4 and 5, extending up to tens of GeV, produces a very good fit to the  observed $\gamma$ rays from solar flares} \citep[][]{LATsolarCat}. \hld{Thus,} the lower energy limit introduced by \citet[][]{PM2023} into the code for a VSSTW search by \citet[][]{PMV2021} is well suited for a blind search for solar flares. \hld{Gamma rays} \hlz{above 10 GeV} \hld{by contrast} {are detected by \textit{Fermi}-LAT from GRBs, for example  a 95 GeV photon from GRB 130427A and a 99 GeV photon from GRB 221009A} \citep[][]{FermiGRB130727A, GCN5}, and also from the Sun during the solar minimum in 2008-2009 (e.g. six \hld{photons} above 100 GeV reported by \citealt[][]{Linden2018}). \hld{To include $\gamma$ rays in the range of tens to hundreds of GeV,} we selected the upper energy limit of 500 GeV.

We downloaded the \fermilat{} Pass 8 (P8R3) \texttt{SOURCE} class (evtype=128) data, collected from 2008 August 4 to 2024 January 9,\LEt{***see note  23 re dates.  } from the Fermi Science Support Center.
\hld{The inclusion of} \hdm{four years} of additional data to the data set \hbb{analysed} by \citet[][]{PMV2021} \hld{allowed us to conduct a search for more recent $\gamma$-ray transients.}
We performed \hld{the data reduction and exposure calculations} using the \texttt{FERMITOOLS} v1.2.23 package. We rejected $\gamma$ rays with
zenith angles larger than $90^{\circ}$ to reduce contamination by
albedo \grays{} from the Earth and applied the recommended cuts on
the data quality (DATA\_QUAL$>0$ \&\& LAT\_CONFIG$==1$). We binned
the data into time intervals of \hlz{six hours} and in four energy bands: 
 0.5-0.8 GeV, 0.8-2.0 GeV, 2.0-5.0 GeV, and 5.0-500.0 GeV. 
We attributed one degree of freedom to every energy band when comparing a likelihood ratio test statistic to a $\chi^2$ distribution.
\hbb{Treatment} of these four energy bands provided us with an analysis independent on the photon index \hlz{and suitable both for GRBs with harder spectra and solar flares with softer spectra}.
We further binned the \fermilat{} $\gamma$ rays using the HEALPix package into a map
of resolution $N_\mathrm{side}$ = 512 in Galactic coordinates with
RING pixel ordering. With these settings, the total number of
pixels is equal to $12\times512^2$=3145728 and the area of each pixel is
$4\pi\times(180/\pi)^2/(12\times512^2)$=0.0131 deg$^2$, given \hld{the fact} that HEALPix subdivides the sphere into 12 equal-area base pixels at the lowest resolution. To compute the exposure, we used the standard tools
\texttt{gtltcube} and \texttt{gtexpcube2}, \hld{which are part of} \texttt{FERMITOOLS}. To correct the livetime
for the zenith angle cut, we used the `zmax' option \hld{in} \texttt{gtltcube}.
For this analysis, we selected \hlz{196608} positions whose centres coincide with those of the HEALPix grid of resolution $N_\mathrm{side}$ = \hlz{128} in Galactic coordinates with RING pixel ordering.
\hld{To search for transient signals at different positions on the sky,} we \hld{grouped together $\gamma$ rays} within a 0\fdg5 \hld{distance from} each of these \hlz{196608} positions. \hlr{This distance is comparable to the HEALPix mean spacing, 0\fdg46, for $N_\mathrm{side}$ = 128.  The high-resolution map with $N_\mathrm{side}$ = 512 resulted in more circular shapes for these groups.}
The 0\fdg5 radius aperture \hld{covers} an area two times larger than \LEt{***twice
as large as } the aperture used in \citet[][]{PMV2021}. \hbb{This} \hlz{increases the number of $\gamma$ rays from the potential source}, but \hld{is} small \hld{enough} to suppress contamination \hld{caused by the PSF wings of highly variable} blazars or other neighbouring sources, especially in the two lowest energy bands.

\section{Results}

We present the results in Table \ref{T1}, which contains the list of \hld{short} $\gamma$-ray transients detected \hld{with} the VSSTW search method, \hld{but excluding transients corresponding to quiescent solar emission or to AGNs}. The names of transients are in the format FST YYMMDD,\footnote{Due to the choice of six-hour time intervals, transients that happened near midnight in UTC, including, for example, GRB 080916C, FLSF 2012-03-07, FLSF 2014-02-25, and GRB 170916A, can have different dates in their FST identifier and associated source name.} where FST stands for \textit{Fermi}-LAT short transient. \hlw{We tabulated the} global significance \hlw{taking into account the trial factor for the weak detection criterion}.
\hld{The number of} the detected GRBs and solar flares listed in this table is much higher than that previously reported by \citet[][]{PMV2021}. If two or more signals belonging to neighbouring spatial pixels result in detections happening close in time, then we list the signal of the highest significance among them.
In addition, we provided evidence for a new transient signal, which we tentatively associated  with a GRB candidate, GRB 180112, and \hlr{reported $\gamma$-ray emission from} a new solar flare that occurred on 2023 December 31.

\begin{table*}
\centering \caption{ Transient \gray{} signals obtained
from the performed VSSTW analysis. The second and third
columns show the right ascension and the declination of a transient \gray{} source.
The fourth and fifth columns show the start date and the length
of a high-\gray{}-flux state. The six and seventh columns show the local and
global significances at which the high flux state is detected.
The eighth and ninth
columns show the name and class of a \gray{} source associated with a transient signal. }
\begin{tabular}{ | c | c | c | c | c | c | c | c | c |}
\hline
Identifier &  R.A. & Dec.  & \#Slice/ &  Length  & Local. &  Global & Assoc. source & Class  \\
 & (deg)       &  (deg)     &  \# interval  &   (6 hours)  & signif. ($\sigma$)  & signif. ($\sigma$)   & &        \\
& &  &  &   &  &  & & \\
\hline
FST 080915 & 119.9 & -57.9 & 1/169 & 1 & 17.0 & $>5$ & GRB 080916C & GRB \\
FST 090322 & 190.8 & 17.4 & \hlw{2}/322 & 1 & 6.9 & 3.4 & GRB 090323 & GRB \\
FST 090328 & 90.9 & -42.1 & 2/343 & 2 & 10.7 & $>5$ & GRB 090328 & GRB \\
FST 090509 & 333.6 & -26.7 & 2/514 & 1 & 18.6 & $>5$ & GRB 090510A & GRB \\
FST 090902 & 264.1 & 27.4 & 2/976 & 1 & 9.0 & $>5$ & GRB 090902B & GRB \\
FST 090926 & 353.5 & -66.4 & 2/1071 & 1 & 17.3 & $>5$ & GRB 090926A & GRB \\
FST 091003 & 251.3 & 36.7 & 2/1099 & 1 & 6.9 & 3.4 & GRB 091003 & GRB \\
FST 100310 & 315.5 & 46.0 & 3/233 & 4 & 7.9 & $>5$ & V407 Cygni & nova \\
FST 100413 & 192.2 & 8.4 & 3/370 & 1 & 7.8 & 4.9 & GRB 100414A & GRB \\
FST 110307 & 347.9 & -5.4 & 4/180 & 3 & 10.7 & $>5$ & FLSF 2011-03-07 & solar flare \\
FST 110416 & 83.5 & 21.9 & 4/338 & 4 & 21.8 & $>5$ & Crab flare & pwn \\
FST 110906 & 165.4 & 6.3 & 4/914 & 1 & 10.9 & $>5$ & FLSF 2011-09-06 & solar flare \\
FST 120306 & 347.7 & -5.1 & 5/142 & 3 & 54.6 & $>5$ & FLSF 2012-03-07 & solar flare \\
FST 120624 & 170.9 & 8.8 & 5/582 & 1 & 7.6 & 4.6 & GRB 120624B & GRB \\
FST 120707 & 105.9 & 22.7 & 5/629.5 & 4 & 7.0 & 3.7 & FLSF 2012-07-06 & solar flare \\
FST 120710 & 94.3 & -70.8 & 5/646 & 1 & 7.1 & 3.8 & GRB 120711A & GRB \\
FST 130327 & 218.1 & -69.9 & 6/183 & 1 & 10.3 & $>5$ & GRB 130327B & GRB \\
FST 130427 & 173.0 & 27.5 & 6/307 & 1 & 30.9 & $>5$ & GRB 130427A & GRB \\
FST 130821 & 314.3 & -11.7 & 6/773 & 1 & 6.8 & 3.2 & GRB 130821A & GRB \\
FST 131011 & 196.4 & -7.2 & 6/975 & 1 & 15.3 & $>5$ & FLSF 2013-10-11 & solar flare \\
FST 131231 & 10.8 & -1.8 & 6/1299 & 1 & 10.8 & $>5$ & GRB 131231A & GRB \\
FST 140206 & 315.4 & -8.6 & 6/1447 & 1 & 10.4 & $>5$ & GRB 140206B & GRB \\
FST 140224 & 338.4 & -9.1 & 7/22 & 1 & 42.1 & $>5$ & FLSF 2014-02-25 & \hlw{solar flare} \\
FST 140810 & 119.1 & 27.8 & 7/689 & 2 & 7.6 & 4.6 & GRB 140810A & GRB \\
FST 140901 & 160.6 & 8.0 & 7/776 & 1 & 23.7 & $>5$ & FLSF \hlw{2014-09-01} & solar flare \\
FST 140928 & 44.3 & -55.7 & 7/884 & 1 & 6.8 & 3.2 & GRB 140928A & GRB \\
FST 141207 & 160.0 & 3.7 & 7/1165 & 1 & 8.3 & $>5$ & GRB 141207A & GRB \\
FST 150523 & 115.2 & -45.6 & 8/331 & 1 & 7.3 & 4.1 & GRB 150523A & GRB \\
FST 150527 & 117.3 & -51.4 & 8/471 & 1 & 10.7 & $>5$ & GRB 150627A & GRB \\
FST 160509 & 310.9 & 76.3 & 9/239 & 3 & 7.2 & 4.0 & GRB 160509A & GRB \\
FST 160623 & 315.4 & 42.4 & 9/419 & 1 & 9.1 & $>5$ & GRB 160623A & GRB \\
FST 160625 & 308.8 & 6.8 & 9/430 & 1 & 14.5 & $>5$ & GRB 160625B & GRB \\
FST 160821 & 171.2 & 42.3 & 9/657 & 1 & 7.9 & $>5$ & GRB 160821A & GRB \\
FST 161107 & 275.0 & -28.3 & 9/970 & 4 & 8.0 & $>5$ & V5856 Sgr & nova \\
FST 170214 & 256.0 & -2.1 & 9/1364 & 1 & 10.2 & $>5$ & GRB 170214A & GRB \\
FST 170915 & 203.9 & -46.8 & 10/678 & 1 & 8.8 & $>5$ & GRB 170906A & GRB \\
FST 170910 & 168.9 & 5.0 & 10/697 & 1 & 52.3 & $>5$ & FLSF 2017-09-10 & solar flare \\
FST 171010 & 66.6 & -10.3 & 10/817 & 1 & 8.0 & $>5$ & GRB 171010A & GRB \\
FST 171117 & 194.7 & -63.8 & 10/962 & 1 & 8.2 & $>5$ & PSR B1259-63 & $\gamma$-ray binary \\
FST 180112 & 43.8 & 76.5 & 10/1193 & 1 & 7.1 & 3.8 & new & GRB candidate \\
FST 180210 & 1.9 & 18.2 & 10/1308 & 1 & 7.7 & 4.8 & GRB 180210A & GRB \\
FST 180414 & 158.8 & -59.7 & 11/59 & 4 & 14.3 & $>5$ & V906 Carinae & nova \\
FST 180720 & 0.1 & -2.7 & 11/448 & 1 & 7.7 & 4.8 & GRB 180720B & GRB \\
FST 181225 & 347.9 & -9.9 & 11/1079.5 & 1 & 7.0 & 3.7 & GRB 181225A & GRB \\
FST 200713 & 59.5 & -54.4 & 13/346 & 4 & 7.9 & $>5$ & YZ Reticuli & nova \\
FST 210809 & 267.4 & -6.7 & 14/412 & 4 & 19.1 & $>5$ & RS Ophiuchi & nova \\
FST 211023 & 73.2 & 85.1 & 14/712 & 1 & 7.3 & 4.1 & GRB 211023A & GRB \\
FST 220408 & 92.7 & -50.9 & 14/1379 & 1 & 8.2 & $>5$ & GRB 220408B & GRB \\
FST 221017 & 156.8 & 15.9 & 15/161 & 1 & 8.9 & $>5$ & GRB 220617A & GRB \\
FST 220627 & 201.0 & -32.7 & 15/201 & 1 & 7.5 & 4.5 & GRB 220627A & GRB \\
FST 221009 & 288.5 & 19.8 & 15/616 & 3 & 14.6 & $>5$ & GRB 221009A & GRB \\
FST 221023 & 230.3 & 15.0 & 15/673 & 1 & 24.9 & $>5$ & GRB 221023A & GRB \\
FST 230812 & 248.9 & 47.7 & 16/344 & 1 & 9.0 & $>5$ & GRB 230812B & GRB \\
FST 231231 & 280.9 & -23.2 & 16/908 & 2 & 7.7 & 4.8 & new & solar flare \\
\hline
\end{tabular}
\label{T1}
\end{table*}

\begin{figure*}
\centering
  \begin{tabular}{@{}cc@{}}
    \includegraphics[angle=0, width=.9\textwidth]{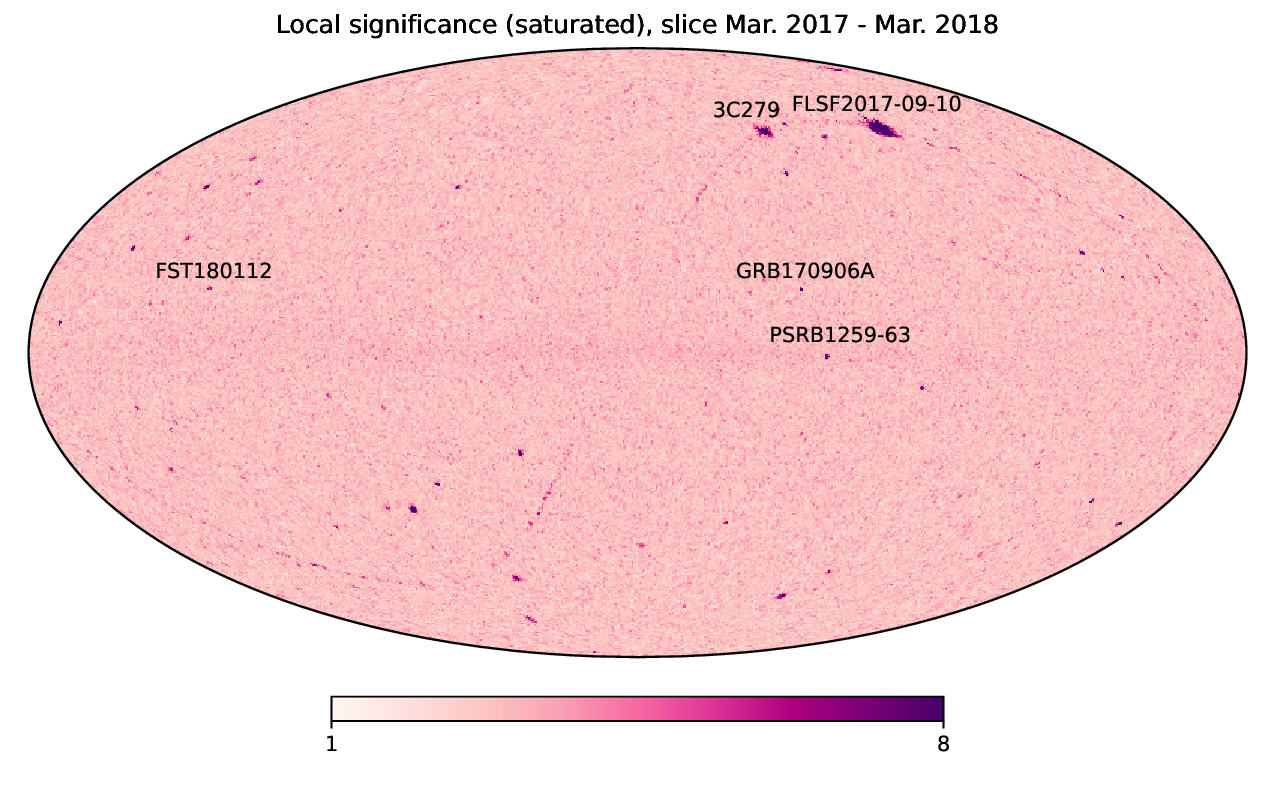}
   \end{tabular}
  \caption{Computed local significance map in $\sigma$. The scale is saturated at a statistical level of 8$\sigma$. The highest significance is 52$\sigma$ for a solar flare, FLSF 17-09-10. The new transient reported in this paper, FST 180112, is shown in this figure along with a blazar, 3C 279; a $\gamma$-ray binary, PSR B1259-63; and GRB 170906A.\LEt{***have I grouped correctly? 3 objects, yes? or 5? }} \label{F1}
\end{figure*}

Figure \ref{F1} shows the local significance map for the ninth slice, based on the data between 2017-03 and 2018-03. The significance maps obtained from the VSSTW analysis applied to each of the analysed slices of data are publicly available in the fits format on Zenodo\footnote{\burl{https://zenodo.org/records/11217341}}.\LEt{*** remove or update this  last sentence; see also note 36}

To differentiate AGNs from other sources, we used both the spatial and temporal information.
We classified a short transient as an AGN if it is located a distance of less than $0\fdg5$ from any gamma-ray source identified or associated with AGNs in the fourth Fermi-LAT catalogue \citep[4FGL;][]{4FGLcat1, 4FGLcat2} and lasts for more than two consecutive six-hour intervals. In the 4FGL catalogue AGNs include mainly BL Lacertae objects, flat spectrum radio quasars, blazar candidates of uncertain type, and radio galaxies.
\hlr{Identifications, such as those used in the 4FGL catalogue, require a correlated flux variability observed at other wavelengths, while associations are positional coincidences that are statistically unlikely to be due to chance alignment between known and candidate $\gamma$-ray producing objects. We do not make a distinction between associations and identifications in this paper.} \hlru{The number of AGNs} with detected flares, that fulfil the  strong criterion, \hlr{defined in Sect.} \ref{Sect2p2}, is 137. \hlr{We found that 24\% (76\%) of these 137 AGNs are located near $\gamma$-ray sources identified as (associated with) an AGN in the 4FGL catalogue. We found that most of these identified or associated AGNs (82\% and 69\%, respectively) correspond to flat spectrum radio quasars.}
Several of these AGNs reveal multiple flares. The sample of AGNs showing at least 5 flares includes PKS 1510-089 (11 flares), 3C 454.3 (9 flares), CTA 102 (7 flares), 3C 279 (6 flares; see Figure \ref{F1}), PKS 1830-211 (5 flares), BL Lac (5 flares), and PKS 0903-57 (5 flares). 

\hlru{The total number of AGNs with detected flares, that fulfil either of the two criteria, defined in Sect.} \ref{Sect2p2}, \hlru{is 178.}\LEt{***either of the two criteria (=there are only 2 criteria, only 1 is needed); two of the criteria (=there are more than 2 criteria, but   only 2 are needed)  }
Most of the detected flares \hlru{($\simeq 90$\%)} from AGNs indeed last for more than two six-hour intervals and both these spatial and temporal conditions are satisfied, allowing easy \hlru{association}. However, there are AGN flares short enough to \hld{lie inside} two six-hour intervals or even one six-hour interval. Among these shortest AGN flares are flares from PKS 1622-253, BL Lac, 1H 0323+342, PMN 2345-1555, S4 0954+65, 4C +31.03, CTA 102, PKS 0310+013, MG2 J071354+1934, 4C +38.41, PKS 2345-16, PKS 1313-333, and PKS 0501-36. 
All these shortest AGN-related transients have sky positions within a $0\fdg33$ radius around AGN positions, allowing accurate \hlru{association}. \hlru{This   suggests that the approach for differentiating GRBs and solar events from AGN flares works well.} In addition to AGNs, there is a short transient with a duration of two six-hour intervals in the direction of an unclassified 4FGL source, 4FGL J0809.8+0507. \hll{The other unclassified 4FGL sources detected in this VSSTW search are 4FGL J0009.2+6847 and 4FGL J1346.5+5330.} \hlr{By the time the most recent 4FGL-DR4 catalogue} \citep[][]{DR4cat} \hlr{was released, our results had been finalised for the first 15 slices. In addition, we  checked and found that among these three sources, 4FGL J0809.8-0507 is associated with an AGN (CRATES J080839+051459), 4FGL J0009.2+6847 is replaced with 4FGL J0008.4+6837 and is associated with an AGN (TXS 0005+683), and 4FGL J1346.5+5330 is associated with an AGN (1RXS J134545.1+533300) in the 4FGL-DR4 catalogue.}

Below we describe non-AGN transient $\gamma$-ray sources detected with the VSSTW algorithm. The detected transients belong to different classes of sources, including GRBs, solar flares, novae, pulsar wind nebulae, and $\gamma$-ray binaries. In addition, there are \hll{numerous} transient events due to the strong $\gamma$-ray emission from the quiescent Sun during the solar minimum as well as \hbb{three} unidentified sources present in the 4FGL catalogue. \hlr{To identify non-AGN transient $\gamma$-ray sources, we used both their celestial coordinates and their flaring activity times. For GRBs and solar flares, we searched for counterparts in the second catalogue of LAT-detected GRBs} \citep[2FLGC;][]{LATgrbCat} \hlr{and in the first} \textit{Fermi}\hlr{-LAT solar flare (FLSF) catalogue}
\citep[][]{LATsolarCat}, \hlr{respectively. For quiescent solar emission, we compared the celestial coordinates of transients with the locations of the Sun, computed by means of the} \texttt{get\_sun} \hlr{routine from the} \texttt{PYTHON} module, \texttt{astropy.coordinates},\footnote{\burl{https://astropy.org}} \hlr{at the given times.}

\subsection{GRBs}

The first Fermi-LAT GRB catalogue was published in 2013 \citep[1FLGC;][]{1stCatGRBs}. It is a compilation of the \hbb{35 GRBs: 30 long-duration ($>$ 2 s)} and 5 short-duration ($<$ 2 s) GRBs, detected in the period 2008 August through 2011 July. Of these GRBs, 28 were detected with standard analysis techniques at energies above 100 MeV, while 7 GRBs were detected only $<$ 100 MeV. The second catalogue of LAT-detected GRBs \citep[2FLGC;][]{LATgrbCat}, covering the first ten years of operations, from 2008 August 4 to 2018 August 4, has a total of 186 GRBs; of these, 169 are detected above 100 MeV. This ten-year \hlw{catalogue} includes 169 long GRBs and 17 short GRBs. During this time, the \textit{Fermi} \hbb{Gamma-ray Burst Monitor} (GBM) triggered on 2357 GRBs, approximately half of which were in the field of view of the \textit{Fermi}-LAT at the time of trigger. In the 1FLGC, the longest duration reported was $>$ 800 s for GRB 090902B. In the 2FLGC, many GRBs have durations of order 1000 s, with the longest duration being 35 ks (GRB 160623A). \hll{Thus, the choice of six-hour intervals is proper for} a \hbb{search} for GRBs in \textit{Fermi}-LAT data.

The list of GRBs detected by means of the previous VSSTW analysis and taken from \citet[][]{PMV2021} includes GRB 080916C, GRB 090510A, GRB 090926A, GRB 130427A, GRB 131231A, GRB 140206B, GRB 160623A, GRB 160625B, and GRB 171010A. The GRBs from this list are present in the 2FLGC catalogue. These nine GRBs have a high \textit{Fermi}-LAT fluence in the 100 MeV–100 GeV energy range \citep[see Table 4; ][]{LATgrbCat}. Among them, the highest fluence is $(45\pm4)\times10^{-5}$ erg$~$cm$^{-2}$ for GRB 130427A. The fluences of the other eight \hbb{GRBs} are $(5.6\pm0.8)\times10^{-5}$ erg$~$cm$^{-2}$ for GRB 080916C, $(3.6\pm0.6)\times10^{-5}$ erg$~$cm$^{-2}$ for GRB 090510A, $(15\pm2)\times10^{-5}$ erg$~$cm$^{-2}$ for GRB 090926A, $(4\pm2)\times10^{-5}$ erg$~$cm$^{-2}$ for GRB 131231A, $(2.3\pm0.6)\times10^{-5}$ erg$~$cm$^{-2}$ for GRB 140206B, $(11\pm3)\times10^{-5}$ erg$~$cm$^{-2}$ for GRB 160823A, $(2.7\pm0.3)\times10^{-5}$ erg$~$cm$^{-2}$ for GRB 160825B, and $(1.1\pm0.3)\times10^{-5}$ erg$~$cm$^{-2}$ for GRB 171010A. Although they are among the highest fluence GRBs, there are other GRBs with high fluences in Table 4 \citep[see][]{LATgrbCat}, including GRB 090902B ($(9\pm1)\times10^{-5}$ erg$~$cm$^{-2}$), GRB 090323 ($(3.4\pm0.9)\times10^{-5}$ erg$~$cm$^{-2}$), and GRB 150627A ($(4\pm0.1)\times10^{-5}$ erg$~$cm$^{-2}$). The presence of GRB 090902B in the 2FLGC catalogue, but not in the list from \citet[][]{PMV2021} was one of the motivations to perform a new VSSTW search using six-hour time intervals.

The VSSTW analysis \hbb{presented} in this paper resulted in detections of \hdm{37} GRBs. Among them, there are 30 GRBs that occurred between 2008 August 4 and 2020 January 30. Thus, the number of GRBs detected with a VSSTW technique increased more than three times compared with the previous search performed by \citet[][]{PMV2021}. The nine GRBs listed in \citet[][]{PMV2021} \hll{are among the detected GRBs} in this analysis. GRB 090902B, GRB 090323, and GRB 150627A are among the mentioned 30 GRBs. The fluences \citep[][]{LATgrbCat} of 17 GRBs from the list obtained by removing the 9 GRBs reported in \citet[][]{PMV2021} along with GRB 090902B, GRB 090323, GRB 150627A, are mostly in the range of ($1\times10^{-5}$, $3\times10^{-5}$) erg$~$cm$^{-2}$. However, the fluences of GRB 091003 and GRB 160509A are only $(0.8\pm0.2)\times10^{-5}$ erg$~$cm$^{-2}$ and $(0.5\pm0.3)\times10^{-5}$ erg$~$cm$^{-2}$, respectively. We note that these 2 GRBs are among the 10 GRBs (out of \hdm{37} GRBs) that satisfy the weak detection criterion, the other \hll{satisfy} the  strong  detection criterion. \hlw{Thus, this VSSTW analysis is sensitive to GRBs with lower fluences than those detected in the previous VSSTW analysis.}

In addition to the GRBs from the the 2FLGC catalogue, the VSSTW analysis resulted in detections of \hdm{eight GRBs}, which occurred between 2018 August 4 and \hdm{2024 January 9}. These \hdm{eight GRBs} include GRB 181225A, GRB 211023A, GRB 220408B, GRB 220617A, GRB 220627A, GRB 221009A, GRB 221023A, \hdm{and GRB 230812B}. \hll{The first was revealed by an analysis of the data set with the three-hour shift.}
All eight of these GRBs had previously been reported in GRB Coordinates Network (GCN) circulars, \#23561 \citep[GRB 181225A;][]{GCN0} \#30961 \citep[GRB 211023A;][]{GCN1}, \#31896 \citep[GRB 220408B;][]{GCN2}, \#32212 \citep[GRB 220617A;][]{GCN3}, \#32283 \citep[GRB 220627A;][]{GCN4}, \#32658 \citep[GRB 221009A;][]{GCN5}, \#32831 \citep[GRB 221023A;][]{GCN6}, \hdm{and \#34392} \citep[GRB 230812B;][]{GCN7}.

\subsection{GRB candidate: FST 180112}
\label{fst180112}

\begin{figure}
\centering
  \begin{tabular}{@{}cc@{}}
    \includegraphics[angle=0, width=.45\textwidth]{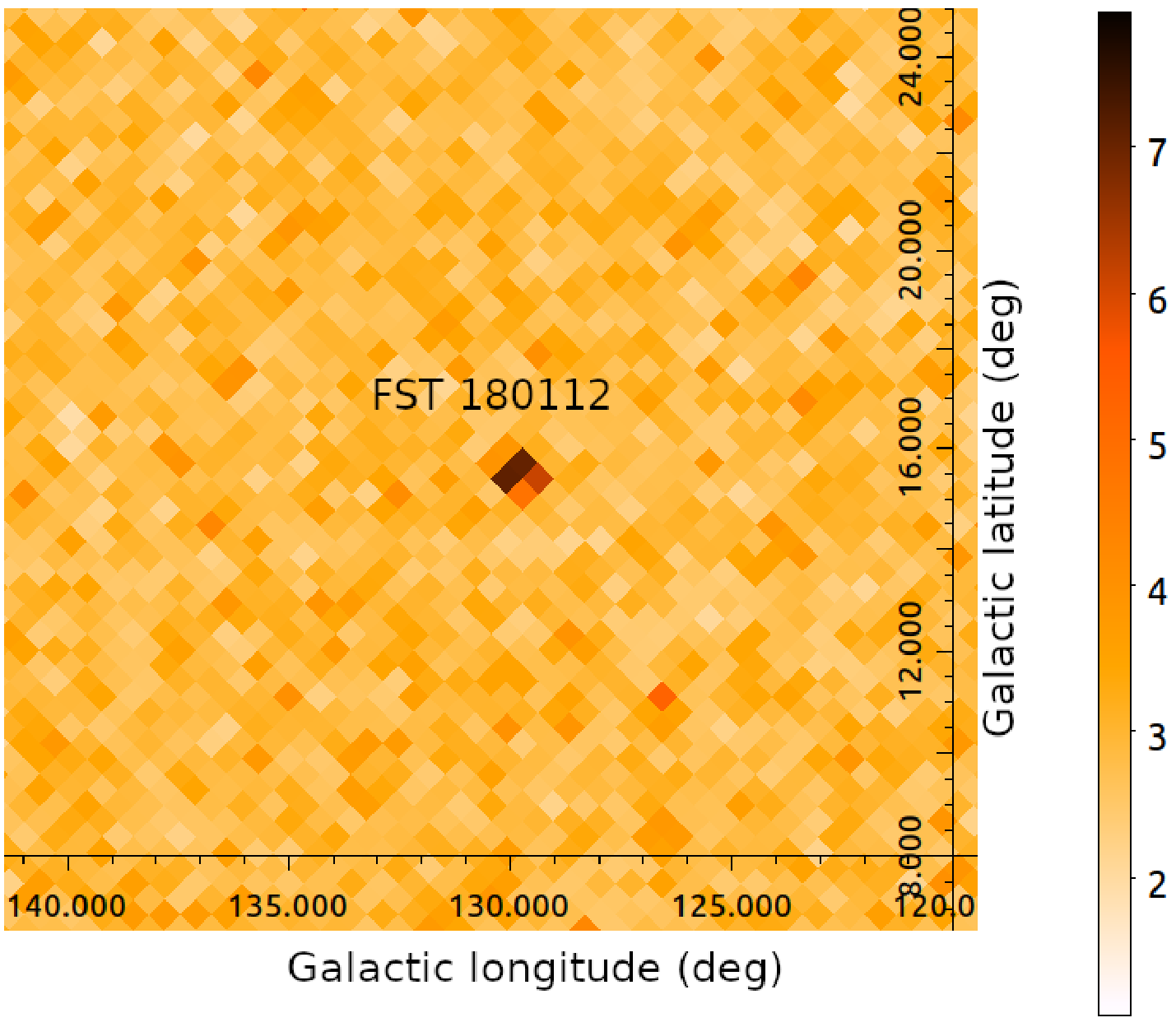}
   \end{tabular}
  \caption{Computed local significance map in $\sigma$ centred on the position of FST 180112.} \label{F2}
\end{figure}

For the sake of robustness, we report new transient sources \hlr{of GeV $\gamma$ rays}, which met the weak detection criterion in both the analysed samples based on the default binning and the one shifted in time
by three hours. The term `new' in this paragraph is applied to GeV $\gamma$-ray sources with unknown \hbb{counterparts} in the Fermi source catalogues or not yet reported via GCN circulars \hdm{or other publications}. This \hlr{detection} criterion allows us to eliminate possible spurious signals in disproportionally short exposure intervals and makes new transient sources reliable. The disproportionally short exposure intervals may arise due to the \hbb{overlap} between the six-hour binning and visibility of the source's position with \textit{Fermi}-LAT. We checked and found that there is only a single event due to the disproportionally short exposure effect \hbb{present} in the sample shifted in time. That event occurred in 2021 November, and  \hbb{as it is}  based only on three $\gamma$ rays, was discarded. The VSSTW analysis resulted in only one new transient source, satisfying the robustness criterion, \hdm{in addition to the solar flare on 2023 December 31}. This transient occurred on 2018 January 12 and its $\gamma$-ray emission is confined within one six-hour interval. The HEALPix pixel corresponding to this transient source has coordinates\LEt{***make sure your abbreviations are all the same. In the tables you use R.A. and Dec. Either is  fine, but please be consistent. (see also note 32) } (R.A., Dec)=(43\fdg83, 76\fdg52). Figure \ref{F2} shows a zoomed-in image of a  local significance map centred on FST 180112.

To study the origin of the  $\gamma$-ray emission \hbb{that} happened on 2018 January 12, we selected $\gamma$ rays recorded by \textit{Fermi}-LAT \hlm{during the six-hour interval covering this transient. This selection revealed a cluster of five $\gamma$ rays contributing to this transient event. All  five of the  $\gamma$ rays are within a 0\fdg5 radius around the corresponding HEALPix pixel. The coordinates and arrival times of these $\gamma$ rays are listed in Table} \ref{Tab12thJan}. \hlm{These five $\gamma$ rays were recorded within a time interval of 11 minutes.}

\begin{table}[t!]
\centering \caption{Five $\gamma$ rays associated with the new transient source, FST 180112. The first and second columns show the right ascension and the declination of these $\gamma$ rays, while the third and fourth columns show their \hbb{arrival times and energies}.}
\begin{tabular}{ | c | c | c | c | c |}
\hline
\# &  R.A. & Dec.  &  Arrival time (MET) & Energy \\
 & (deg)       &  (deg)         &  (s)   & GeV    \\
& &  &  & \\
\hline
I & 44.321 & 76.655 & 537468146 & 1.82 \\
II & 42.613 & 76.815 & 537468224 & 3.88 \\
III & 43.246 & 76.849 & 537468253 & 4.79 \\
IV & 43.692 & 76.846 & 537468418 & 0.91 \\
V & 42.552 & 76.795 & 537468797 & 3.40 \\
\hline
\end{tabular}
\label{Tab12thJan}
\end{table}

Given that the duration of FST 180112 is short, its position is more than 100$^{\circ}$ away from the Sun, it is $15^{\circ}$ above the Galactic plane, and it is not associated with 4FGL sources, this transient is likely associated with either a GRB or an AGN flare. As we stated above, most of the AGN flares revealed with this VSSTW analysis has a duration longer than six hours. Thus, FST 180112 is more likely associated with a GRB. We searched for a GRB that occurred on 2018 January 12 in the internet resources, including the  SPI-ACS/INTEGRAL,\footnote{\burl{https://gcn.gsfc.nasa.gov/integral_spiacs.html}} \hlw{AstroSat CZTI},\footnote{\burl{http://astrosat.iucaa.in/czti/?q=grb}} Konus-Wind,\footnote{\burl{http://www.ioffe.ru/LEA/kw/triggers/2018/index.html}} and CALET-GBM\footnote{\url{https://gcn.gsfc.nasa.gov/calet_triggers.html}} \hlr{webpages}. The source `possible GRB 180112' was seen on 2018 January 12 with SPI-ACS/INTEGRAL at 16:30:39 UTC,\footnote{\burl{https://gcn.gsfc.nasa.gov/other/7990.integral_spiacs}} \hlw{AstroSAT CZTI} at \hlw{16:30:38 UTC,}\footnote{\burl{http://web.iucaa.in/~astrosat/czti\_grb/GRB180112A/AS1CZT_GRB180112A\_10SECBIN\_lightcurves.pdf}} Konus-Wind at 16:29:11 UTC,\footnote{\burl{http://www.ioffe.ru/LEA/kw/triggers/2018/kw20180112\_59351.html}} and CALET-GBM at \hlw{16:29:08} UTC.\footnote{\burl{https://gcn.gsfc.nasa.gov/other/1199809543.calet}} \hlw{The light curves based on the  Konus-Wind data and the CALET-GBM data look identical and show three separate peaks in the count rates, while the light curves based on the  SPI-ACS/INTEGRAL data and the  AstroSat data begin with the second peak. This explains the later start time of the transient recorded by SPI-ACS/INTEGRAL and AstroSat.} The probability \LEt{***probability? also below} that FST 180112 and the event simultaneously observed with SPI-ACS/INTEGRAL, Konus-Wind, CALET-GBM, \hdm{and AstroSat} is a pure coincidence in time is about 1\%, given the daily GRB detection rate \hlw{recorded} by these instruments. \hlr{This probability is given by} $\simeq f \Delta t$, \hlr{where} $f=1/24$ hour$^{-1}$ \hlr{and the time interval between the signal detected by CALET-GBM and} \textit{Fermi}-LAT, $\Delta t\sim1000$ \hlr{seconds}.

\hdm{No corresponding signal is present in the Fermi GBM Burst Catalogue}\footnote{\burl{https://heasarc.gsfc.nasa.gov/W3Browse/fermi/fermigbrst.html}} \hdm{and the Fermi GBM Trigger Catalogue}.\footnote{\burl{https://heasarc.gsfc.nasa.gov/W3Browse/fermi/fermigtrig.html}} \hlr{The FGST spacecraft was passing through the South Atlantic Anomaly from 16:09:40 UTC to 16:35:14 UTC on 2018 January 12. Thus, the} \textit{Fermi} \hlr{GBM detectors were disabled and recorded no signal between 16:29:08 UTC and 16:30:39 UTC.} The first of the five $\gamma$ rays reported in Table \ref{Tab12thJan} was recorded by \textit{Fermi}-LAT on 16:42:21 UT.

\hdm{A uniform all-sky Candidate Gamma-Ray Blazar Survey (CGRaBS) selected primarily by flat radio spectra was performed by} \citet[][]{Healey} \hdm{to provide a large catalogue of likely gamma-ray AGNs suitable for identification of high-latitude $\gamma$-ray sources detected with} \textit{Fermi}-LAT. \hdm{We found that the closest source in the CGRaBS catalogue to FST 180112 is CRBABs J0257+7843 and is located 2 degrees from the position of FST 180112, suggesting that the GRB nature of this transient is a good working hypothesis.}

\hdm{We encourage members of the GRB community to perform interplanetary network triangulation of the GRB candidate to localise the signal in space for comparison with the location of FST 180112 to identify its nature in an unambiguous way.}

\subsection{Solar flares}

The first FLSF catalogue covers the 24th solar cycle, \hdm{which lasted from 2008 to 2019}, and  contains 45 solar flares with emission in the $\gamma$-ray energy band (30 MeV - 10 GeV) detected with a significance $>5\sigma$ over the years 2010-2018 \citep[][]{LATsolarCat}. Thirty-seven of these flares exhibit delayed emission beyond the prompt-impulsive hard X-ray (HXR) phase with 21 flares showing delayed emission lasting more than two hours, including the longest extended emission ever detected (20 hours) from the FLSF 2012-03-07. This catalogue includes three flares originating from active regions located behind the limb of the visible solar disk. The analysis by \citet[][]{LATsolarCat} indicates that the solar flare emission above 100 MeV is due to accelerated ions as opposed to HXRs and microwave producing electrons. Two main populations of FLSFs are impulsive-prompt and gradual-delayed.
The impulsive-prompt flares are those whose emission evolution is similar to that of the HXRs, while the emission of delayed FLSF flares rises at the end of the impulsive HXR phase and extends well beyond the end of the HXR emission.

The list of solar flares from \citet[][]{PMV2021} includes three flares on 2012 March 7, 2014 September 1, and 2017 September 10. These three solar flares are present in the FLSF catalogue. These three solar flares have the highest \textit{Fermi}-LAT fluence above 100 MeV among all the solar flares reported in the FLSF catalogue \citep[see Table 1; ][]{LATsolarCat}. Their photon fluences are $33.996 \pm 0.030$ cm$^{-2}$ for FLSF 2012-03-07, $12.1\pm2.3$ cm$^{-2}$ for FLSF 2014-09-01, and $22.2\pm1.6$ cm$^{-2}$ for FLSF 2017-09-10. Another solar flare with such a high fluence from the FLSF catalogue is FLSF 2014-02-25 ($13.95\pm0.18$ cm$^{-2}$). The fluence of FLSF 2014-02-25 is as high as that of FLSF 2014-09-01, but the former was not revealed with the previous VSSTW analysis. Apart from these four solar flares with the highest fluence, the other solar flares from the FLSF catalogue have significantly lower fluences. Thus, the solar flares with the fifth and sixth highest fluences are FLSF 2011-03-07 ($1.076\pm0.029 $ cm$^{-2}$) and FLSF 2017-09-06b ($1.0700\pm0.0022$ cm$^{-2}$). One of the goals of the VSSTW  analysis in this paper is to detect solar flares with significantly lower fluences than those of the three solar flares from \citet[][]{PMV2021}.

The VSSTW analysis in this paper resulted in detections of eight solar flares \hdm{during the 24th solar cycle}, including FLSF 2011-03-07, FLSF 2011-09-06, FLSF 2012-03-07, FLSF 2012-07-06, FLSF 2013-10-11, FLSF 2014-02-25, FLSF 2014-09-01, and FLSF 2017-09-10.
Thus, the number of solar flares \hdm{during the 24th solar cycle} detected with a VSSTW technique increased more than 2.5 times compared with the previous search performed by \citet[][]{PMV2021}. In addition to the three solar flares with the highest fluence from \citet[][]{PMV2021}, the VSSTW analysis also revealed two more solar flares, namely FLSF 2014-02-25 and FLSF 2011-03-07, with fluences higher than $1$ cm$^{-2}$. The other three detected flares, FLSF 2011-09-06, FLSF 2012-07-06, and FLSF 2013-10-11, have fluences of  $0.87\pm0.17$ cm$^{-2}$, $0.100\pm0.021$ cm$^{-2}$, and $0.262\pm0.013$ cm$^{-2}$, respectively \citep[][]{LATsolarCat}. Thus, the VSSTW analysis presented in this paper allowed us to lower a fluence for solar flares detected by means a VSSTW technique from $10$ cm$^{-2}$ to below $1$ cm$^{-2}$.

\begin{figure}
\centering
  \begin{tabular}{@{}cc@{}}
    \includegraphics[angle=0, width=.45\textwidth]{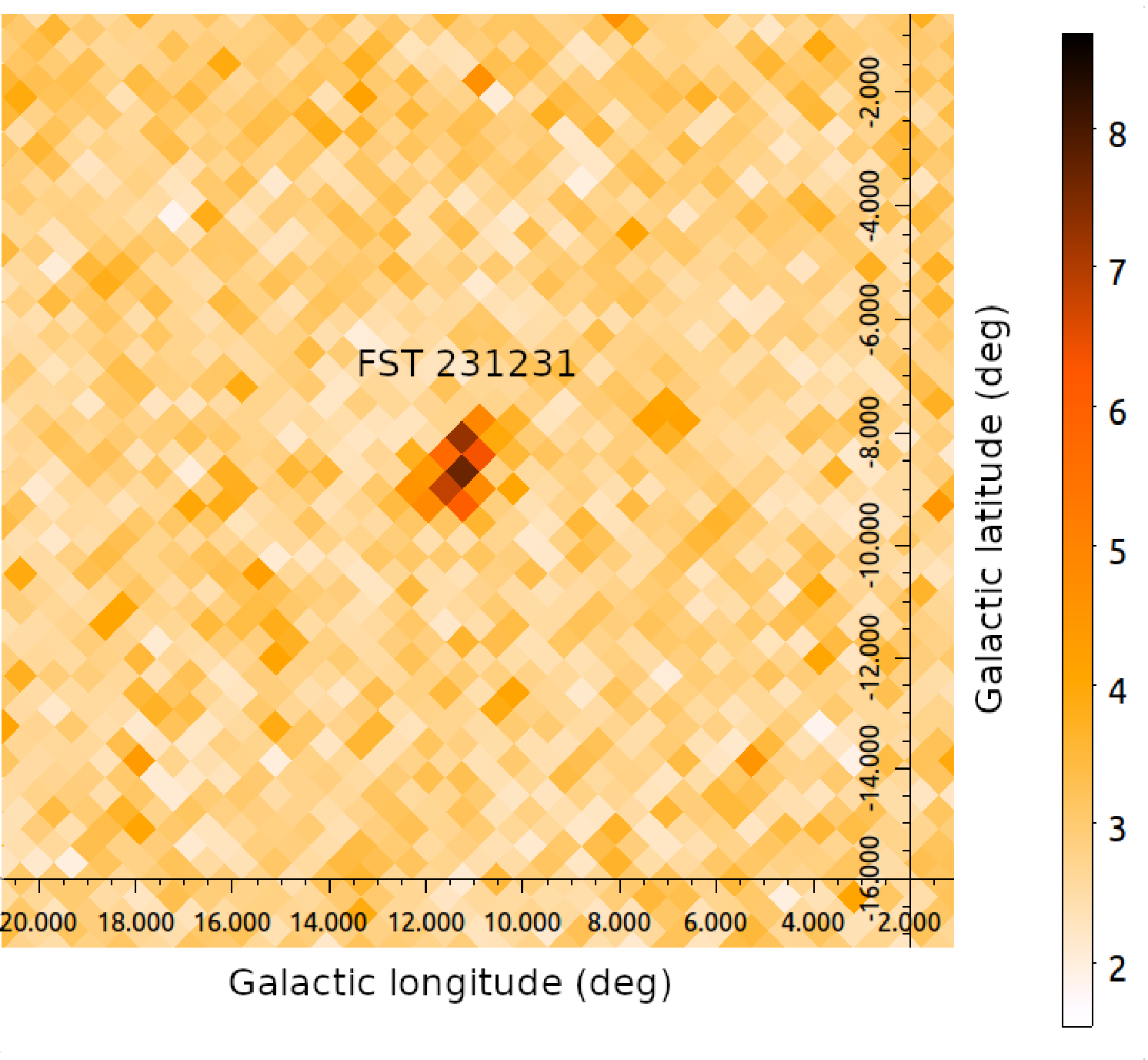}
   \end{tabular}
  \caption{\hdm{Computed local significance map in $\sigma$ centred on the position of FST 231231, corresponding to the solar flare on 2023 December 31.}} \label{F3}
\end{figure}

\hdm{In addition to  the solar flares from} \citet[][]{LATsolarCat}, \hdm{the present VSSTW analysis resulted in a detection of $\gamma$-ray emission from a solar flare on 2023 December 31. The signal based on the VSSTW analysis occurred on 2023 December 31 and its $\gamma$-ray emission is confined within two six-hour intervals. The HEALPix pixel corresponding to this transient source has coordinates (R.A., Dec)=(280\fdg86, -23\fdg18).} Figure \ref{F3} \hdm{shows a local significance map centred on this position. The corresponding solar flare occurred during the 25th solar cycle that started in 2019 and was classified as an X5.0 flare}\footnote{\burl{https://svs.gsfc.nasa.gov/14497/}}. \hdm{It is the largest flare since September 10, 2017 when an X8.2 flare occurred.}

\subsection{Quiescent solar emission}

The detection of high-energy $\gamma$ rays from the quiescent Sun with \textit{Fermi}-LAT was made during the period of low solar activity when the emission induced by cosmic rays is brightest \citep[][]{solar2011}. Because the Sun is moving across the sky, the Sun leaves behind a track on the $\gamma$-ray sky map \citep[see Fig. 1 in][]{occ}. The highly significant detection allowed clear separation of the two components: the point-like emission from the solar disk due to \hlw{cosmic-ray} cascades in the solar atmosphere, and the extended emission from the inverse Compton scattering of \hlw{cosmic-ray} electrons on solar photons in the heliosphere. The fluxes of both \hll{these emission} components must change over the solar cycle due to the change in the heliospheric flux of the Galactic cosmic rays in  anti-correlation with the variations of the solar activity. \hll{The} significant time variation of the $\gamma$-ray flux from the solar disk that anti-correlates with solar activity was indeed detected in the \textit{Fermi}-LAT data \citep[][]{Ng2016, Linden2022}. The Sun-centred maps were used  in the analyses by \citet[][]{Ng2016} \hbb{and} \citet[][]{Linden2022} to study solar flux variations.

The list of \hbb{transients} from \citet[][]{PMV2021} includes only two \hbb{signals} corresponding to the quiescent Sun in 2009 July and 2019 October. In addition to these two transients, they reported a transient signal from the Sun on 2013 \hdm{February} 8. A similar solar transient in 2013 February was later reported in Table 2 of \citet[][]{FermiLTT2021}. In addition to  this transient, \citet[][]{FermiLTT2021} reported 34 transients occurred between 2008 August and 2018 August, and is associated with the quiescent emission from the moving Sun. These 34 transients are shown in Table 2 and Figure 3 in \citet[][]{FermiLTT2021}; 23 of \hlw{these} 34 transients occurred before 2011 March and the  8 most recent of these transients occurred in 2017.

\hlr{The solar track is clearly visible in the significance map shown in Figure} \ref{F1}. \hlr{This significance map is for the ninth slice, which corresponds to the data taken between 2017 March and 2018 March}. \hlr{This track of the Sun is called the ecliptic. The Sun moves eastwards on the celestial sphere and completes one full circle of 360$^{\circ}$ over one year.} \hlr{Solar flare FLSF 2017-09-10 and  Blazar 3C 279 are noticeable on the solar track. Blazar 3C 279} is occulted by the Sun on 8  October every year. \hlw{The solar track in} Figure \ref{F1} covers a large number of pixels \hlr{in the sky}. This number is much larger than 3, which represents the transients corresponding to the quiescent solar emission detected \hlr{from 2017 March to 2018 March and shown in Table} \ref{T1}. However, most of these \hlr{pixels do not meet the weak detection criterion.} The detected transients corresponding to the quiescent solar emission are most likely upward fluctuations of the $\gamma$-ray flux from the solar disk.

The VSSTW analysis \hbb{presented} in this paper resulted in detections of \hlw{35} transients associated with the quiescent emission from the solar disk. \hlr{Consequently, this allowed for the detection of approximately} 20 times more transients \hlr{related to} the \hlw{quiescent} $\gamma$-ray emission from the solar disk than in the previous VSSTW analysis. Twenty of these \hlw{35} solar transients were in 2008-2010, \hbb{of which} 16 were between 2008 August and 2009 September. Different sky-survey profiles have been used on Fermi\footnote{\burl{https://fermi.gsfc.nasa.gov/ssc/observations/types/allsky/}}, and the on-source exposure also varies in time because of the use of different sky-survey profiles. The original rocking profile was 35 degrees north (relative to the local zenith) for one orbit, then 35 degrees south for one orbit. Due to spacecraft thermal considerations Fermi gradually increased the rocking angle until settling on a rocking angle of 50 degrees in 2009 September. The difference in  rocking profile is \hlw{the reason} why the number of transients associated with the solar quiescent emission is higher \hlw{between 2008 August and 2009 September} in the beginning of the 24th solar cycle. \hlw{Four of these 16 transients that occurred before 2009 September are the only ones of the total 35 transients that satisfy the strong detection criterion.} \hbb{Solar} activity was minimal until early 2010 and reached its maximum in 2014 April. The number of solar flares, including M5-M9 and X1-X5, per year was high in \hlw{2011-2015} \citep[e.g. Fig. 20 of][]{LATsolarCat}. Only two transients associated with the solar quiescent emission were detected with the VSSTW analysis between 2011 and 2015. Between 2016 and 2020, the sunspot number was \hbb{decreasing} and the number of transients associated with the solar quiescent emission between 2016 and 2020 was 13. \hlw{It may indicate that} the change in the rate of transients \hbb{between 2011-2015 and 2016-2020 is} associated with quiescent solar emission \hbb{and is} in anti-correlation with the variations of the solar activity.

\subsection{Other transient sources}

In addition to GRBs, solar flares, quiescent solar events, and blazars, there are sources of other classes detected in this VSSTW search.
Most of these transients are novae, including V407 Cygni, V5856 Sagittarii, V906 Carinae, YZ Reticuli, and RS Ophiuchi. The VSSTW analysis shows that the duration of these transients is longer than three six-hour intervals.  This behaviour is similar to that of blazars and should be interpreted that these transients last \hbb{for significantly longer intervals} than those tested in this paper. It is known that $\gamma$-ray emission from novae usually lasts for two weeks \citep[e.g.][]{v407, Franckowiak2018}. The limit of maximum possible duration, used here, thus divides the $\gamma$-ray emission signal from novae into two parts. The longest of these parts will contribute to the background, but given that the background is computed from the one-year interval, this contribution will be  small to substantially increase the background level. Three novae, V407 Cygni, V5856 Sagittarii, and V906 Carinae, are present in the list from \citet[][]{PMV2021} at high significance levels. As for YZ Reticuli, and RS Ophiuchi, they are more recent novae and occurred in 2020 July and 2021 August, \hbb{respectively}. YZ Reticuli is at a high Galactic latitude \citep[][]{FermiLTT2021} and RS Ophiuchi is the brightest \textit{Fermi}-LAT nova \citep[][]{rhoOph}. The higher significances of transient emissions from RS Ophiuchi and V906 Carinae, which are $19.1\sigma$ and 14.3$\sigma$, are due to the fact that these two novae are brighter in $\gamma$ rays than the other three novae \citep[][]{rhoOph}.

Apart from novae, the VSSTW analysis shows transient $\gamma$-ray emission from the Crab nebula and the $\gamma$-ray binary, PSR B1259-63. The transient $\gamma$-ray emission from the Crab nebula is at a significance level of $21.8\sigma$ and corresponds to the exceptional flare of 2011 April \citep[][]{Buehler}. In the previous VSSTW analysis \citep[][]{PMV2021}, this exceptional flare lasted for about one week \citep[][]{Buehler, PMV2021}, its detection in this analysis is explained in the same way as the five novae \hbb{mentioned above}. As for the $\gamma$-ray binary, PSR B1259-63, which has an orbital period of 1237 days, its high-$\gamma$-ray-flux state started in 2011 January, 2017 October, and 2021 March, following the periastron passages \citep[e.g.][]{PM2023}. The transient signal from PSR B1259-63, detected with this VSSTW analysis, occurred \hld{in 2017 November 17} following the periastron passage on \hbb{2017 September 22} and lasted less than six hours (see Figure \ref{F1}). It agrees with the fact that the maximum one-day flux level was 56 days after the periastron passage \citep[][]{Johnson}. Excluding GRBs and solar flares, this flare from PSR B1259-63 is the fastest GeV variability observed in LAT data \citep[][]{Johnson}, and thus the short duration of this flare is expected.

\section{Discussion: FST 180112}

\hlm{In order to clarify whether the presence of five counts near position of FST 180112 recorded in} \hbb{11 minutes} is an ordinary or extraordinary transient event, we performed an additional analysis using a clustering algorithm. The data set \hlw{analysed} here includes the first \hdm{814} weeks of \textit{Fermi}-LAT data and had been recorded between 2008 August 4 and 2024 January 9. Similarly to the VSSTW analysis, we selected $\gamma$ rays above 0.5 GeV. Given that the number of diffuse $\gamma$ rays is high near the Galactic plane, we removed all the $\gamma$-ray events between $-2^{\circ}$ and $+2^{\circ}$ Galactic latitudes. We also excluded the clusters if the tested count is \hbb{within} $1^{\circ}$ of \hbb{a} 4FGL $\gamma$-ray \hbb{source}. The goal of this additional analysis is to search for sets of five counts with a distance not exceeding a given value. For this purpose, we used the \texttt{PYTHON} module, \texttt{sklearn.neighbors}\footnote{\burl{https://scikit-learn.org}} and a nearest neighbours algorithm, \texttt{KDTree}. \hlm{For this clustering analysis, we selected a `Manhattan' distance, which is a sum of absolute values. In  contrast with the clustering algorithms mentioned in Sect.} \ref{CA}, we introduce two spatial coordinates and one time coordinate. The first spatial coordinate is declination, $\mathrm{Dec}$, \hbb{and} the second spatial coordinate is $\mathrm{R.A.}\times\cos\left(\left(\pi/180\right)\times\mathrm{Dec}\right)$, where $\mathrm{R.A.}$ is right ascension.\LEt{***see note 30 re abbreviations } The time coordinate is time measured in seconds and further divided by 600. Thus, one degrees of distance approximately corresponds \hbb{ten minutes} of time. \hlw{Both these spatial coordinates are in  units of degrees.} The selected Manhattan distance is a \hlr{compromise} between the Euclidean distance between two points given their spatial coordinates and the linear distance between two points in time. The Manhattan distance is larger than the Euclidean distance, at most by $\sqrt{2}$, and is a reasonable proxy for estimating a spatiotemporal distance. After finding the spatiotemporal clusters of $\gamma$ rays, we identify them with known $\gamma$-ray sources using the temporal and spatial information.

\begin{table}[t!]
\begin{minipage}{0.47\textwidth}
\centering \caption{\hlr{Gamma-ray sources identified with spatiotemporal clusters of five $\gamma$ rays with distances sorted in ascending order.}}
\begin{tabular}{ | c | c || c | c | }
\hline
Name & Dist. &  Name & Dist. \\
\hline
GRB 090902B & 0.08 & GRB 080916C & 0.46\\
GRB 130427A & 0.10 & GRB 131108A & 0.48\\
GRB 090510A & 0.11 & FLSF 2011-09-06 & 0.48 \\
GRB 090926A & 0.12 & GRB 130327B & 0.52\\
GRB 221023A & 0.15 & GRB 181225A & 0.53\\
FLSF 2014-02-25 & 0.15 & GRB 110731A & 0.55 \\
GRB 170214A & 0.18 & GRB 100116A & 0.56 \\
GRB 160625B & 0.18 & GRB 120711A & 0.60 \\
FLSF 2012-03-07 & 0.21 & GRB 220627A & 0.60 \\
FLSF 2017-09-10 & 0.24 & RS Ophiuchi & 0.63 \\
GRB 220617A & 0.25 & GRB 160821A & 0.67 \\
GRB 221009A & 0.30 & FLSF 2012-06-03 & 0.67 \\
GRB 220408B & 0.30 & GRB 100414A & 0.70 \\
GRB 140402A & 0.31 & GRB 140206B & 0.71 \\
FLSF 2014-09-01 & 0.32 & GRB 190531B & 0.73 \\
FLSF 2013-10-11 & 0.33 & GRB 120624B & 0.73\\
GRB 180720B & 0.36 & \textbf{FST 180112} & 0.76 \\
GRB 141207A & 0.40 & GRB 230812B & 0.80 \\
GRB 160509A & 0.43 & GRB 090328 & 0.89 \\
GRB 131231A & 0.46 & PKS 0903-57 & 0.89 \\
\hline
\end{tabular}
\footnotetext{Notes. For emphasis, FST 180112 is shown in bold.}
\label{TabCluster}
\end{minipage}
\end{table}

\hlr{We} list the spatiotemporal clusters of five $\gamma$ rays with distances sorted in ascending order and 
the $\gamma$-ray sources identified with these clusters \hlr{in Table} \ref{TabCluster}. 
We terminated the cluster analysis when \hbb{it} encountered an AGN flare, which was from PKS 0903-57 and reported in the Astronomer's Telegram \#13604 \citep[][]{ATEL13604}.

There are \hdm{six GRBs} and one FLSF corresponding to clusters of five $\gamma$ rays with distances of  less than 0.89, but not revealed with the VSSTW analysis:\LEt{***if you say "among these" you mean there are others, but you list them all } GRB 100116A, GRB 110731A, FLSF 2012-06-03, GRB 131108, GRB 140402A, GRB 160509A, and GRB 190531B. On the other hand, there are 12 GRBs and \hdm{three} FLSFs detected with the VSSTW analysis, but not revealed with the cluster analysis:  GRB 091003, GRB 090323, FLSF 2011-03-07, FLSF 2012-07-07, GRB 130821A, GRB 140810A, GRB 140928A, GRB 150523A, GRB 150627A, GRB 160623A, GRB 170906A (see Figure \ref{F1}), GRB 171010A, GRB 180210A, GRB 211023A, \hdm{and the solar flare on 2023-12-31}. Although the statistical significance of each of the  clusters of five $\gamma$ rays is not determined, the fact that all the clusters with distances less than that for FST 180112 are identified with known GRBs, FLSFs, and the nova RS Ophiuchi implies that the result obtained from the cluster analysis is reliable. Given that FST 180112 has the smallest distance for not-yet-identified \hbb{signals present in} the the cluster analysis, this transient source deserves an exploration of existing archival multiwavelength observations.

\hlr{The fact that four missions, namely INTEGRAL, AstroSat, Konus-Wind, and CALET, detected a strong GRB (see Sect.} \ref{fst180112}) \hlr{just 1000 seconds before the cluster of GeV $\gamma$ rays suggests that FST 180112 is GRB high-energy afterglow emission as usually detected by} \textit{Fermi}-LAT.   

\section{Outlook}

\hlr{This paper confirms the validity of our previous work and the reliability of the VSSTW technique. The presented VSSTW analysis inherited the energy range, the 0\fdg5 radius aperture, and the} \texttt{SOURCE} \hlr{event class from the previous analysis by} \citet[][]{PM2023}. \hlr{This inheritance was necessary here for a direct comparison of the current and previous results. There is room for some further exploration of $\gamma$-ray transients by means of the VSSTW technique for enabling searches on time intervals much shorter than six hours. The selection of larger data sets can also be useful. To extend the energy range down to the nominal} \textit{Fermi}\hlr{-LAT energy boundary, 100 MeV, for the inclusion of more $\gamma$ rays from GRBs and solar flares, the constant aperture may be changed to an energy-dependent aperture compensating for the increase of the} \textit{Fermi}\hlr{-LAT PSF at low energies. A segmented aperture with radial weights may also be useful to better approximate the PSF. For analyses of timescales $<$200 seconds that benefit from increased photon statistics while tolerating a higher background fraction and broader PSF, the} \texttt{TRANSIENT} \hlr{event class is a good choice and is recommended by the} \textit{Fermi}\hlr{-LAT team. Our} open source code for a VSSTW search \hlr{is available on Zenodo and} provides members of the $\gamma$-ray astronomy community with a tool for future detections of transients.

\section{Summary}

We performed a VSSTW analysis with the purpose of searching for transient $\gamma$-ray signals in the \textit{Fermi}-LAT data. Compared to the previous searches by means of this technique by \citet[][]{PMV2021} and \citet[][]{PM2023}, we used six-hour time intervals instead of seven-day time intervals. The use of shorter time intervals increased \hbb{the} signal-to-noise ratio and allowed more detections of GRBs and solar flares. The detected transient sources, including GRBs and solar flares, are listed in Table \ref{T1}. This refined search increased the number of short transient $\gamma$-ray sources, such as GRBs and solar flares, by a factor of 3 compared to the number reported in the search previously published by \citet[][]{PMV2021}. \hlr{Based on this analysis}, we \hlr{reported} for the first time a $\gamma$-ray signal from the solar flare that occurred on 2023 December 31 during the 25th solar cycle.\LEt{***as written it says you reported it on December 31. or perhaps: Based on this analysis, we reported for the first time a $\gamma$-ray signal from the solar flare that occurred on 2023 December 31 during the 25th solar cycle.} In addition, the performed VSSTW analysis resulted in detections of 37 transient signals associated with the quiescent emission from the solar disk. The distribution of these 37 signals indicates the presence of the solar modulation \hlw{of Galactic cosmic rays}. All these \hbb{show} the efficiency of the VSSTW technique for blind searches for intraday $\gamma$-ray transients.

In addition to  GRBs and transients due to the solar emission, the VSSTW search revealed a number of AGN flares and novae as well as one $\gamma$-ray binary, PSR B1259-63, and the exceptional flare from the Crab nebula. Most of the intraday $\gamma$-ray transients detected with this blind search are identified with known transient sources mainly detected with the analyses that make use of multiwavelength information. Meanwhile, we reported a new transient signal, FST 180112, recorded with \textit{Fermi}-LAT on 2018 January 12. This transient signal \hbb{includes} a cluster of five $\gamma$ rays that all arrived within 11 minutes.\LEt{*** ... rays that all arrived within 11 minutes.  } This $\gamma$-ray transient is \hdm{probably from} a GRB, given its duration and also the presence of a simultaneous lower energy signal detected with SPI-ACS/INTEGRAL, Konus-Wind, CALET-GBM, \hdm{and AstroSat}. FST 180112, revealed with the VSSTW analysis, shows that this technique allows new detections of sources in the archival data although this technique is commonly less sensitive than the traditional (non-blinded) analyses given a trial factor. \hlr{All the transient signals found by means of this VSSTW analysis have source associations.}

\section{Acknowledgements}
\hlr{We are grateful to the referee for the constructive comments that helped us to improve the manuscript.} \hbb{We acknowledge access to high-performance facilities (TIARA cluster and storage) in ASIAA, and the computational facilities of the ALMA Regional Center Taiwan, Academia Sinica, Taiwan.}


\end{document}